\documentclass[11pt,aps,preprint,tightenlines,showpacs,groupadress]{revtex4}
\usepackage{graphicx}
\usepackage{rotating}
\usepackage{amssymb}
\usepackage{mathptmx}

\begin{document}

\title{Path Integral Monte Carlo calculation of momentum distribution in solid $^4$He}

\author{R. Rota$^1$ and J. Boronat$^1$}

\affiliation{
1:Departament de F\'\i sica i Enginyeria Nuclear, Campus Nord B4-B5, Universitat Polit\`ecnica de Catalunya, E-08034 Barcelona, Spain\\
}
\date{22.06.2010}

\begin{abstract}

We perform calculations of the momentum distribution $n(k)$ in solid $^4$He  by means of path integral Monte Carlo methods. We see that, in perfect crystals, $n(k)$ does not depend on temperature $T$ and that is different from the classical Gaussian shape of the Maxwell-Boltzmann distribution, even though these discrepancies decrease when the density of the system increases. In crystals presenting vacancies, we see that for $T \ge 0.75 \, {\rm K}$, $n(k)$ presents the same behavior as in the perfect crystal, but, at lower $T$, it presents a peak when $k \to 0$.

\end{abstract}


\maketitle

\section{Introduction}

The observation, made by Kim and Chan in 2004, of a non classical rotational inertia (NCRI) in a torsional oscillator (TO) containing solid helium\cite{KimChan04Vycor,KimChan04Bulk} has generated, in the last years, a huge interest in the debate about supersolidity. The first theoretical speculations about the supersolid phase, i.e. a phase where crystalline order coexists with superfluidity and Bose-Einstein condensation (BEC), date back almost forty years\cite{AndreevLifshitz69,Chester70,Leggett70}. Nevertheless, we are still far from a complete description of this phenomenon. The evidences of NCRI in solid helium have been confirmed in several TO experiments performed by other groups\cite{Aoki07,Kondo07}, which often gave controversial results. Nowadays, there is an overall agreement of all the data concerning the onset temperature of this phenomenon ($T_c \sim 100 \, \rm{mK}$), but the values of the superfluid density $\rho_s/\rho$ reported so far can vary more than one order of magnitude, according to experimental conditions such as the way the crystal is prepared, its subsequent annealing or the $^3$He concentration\cite{Rittner07}. These discrepancies suggest that the quality of the solid sample plays a very important role in these experiments. So, it is very important to understand the behavior of crystalline defects and to study if they are necessary to have superfluidity in a crystal and if they can be stable in the ground state of this system.

From the theoretical point of view, the presence of BEC in a crystal can be detected by computing the momentum distribution $n(\bf{k})$. The macroscopic occupation of the lowest energy state, in strongly interacting system like $^4$He, appears in $n(\bf{k})$ as a delta-peak for $\bf{k} = 0$ and a divergent behavior $n(k) \sim 1/k$ when $k \to 0$. Equivalently, one can get information about the BEC properties of a quantum system from the asymptotic behavior of the one-body density matrix $\rho_1(\bf{r})$, which is the inverse Fourier transform of $n(\bf{k})$. If at large distances $\rho_1(r)$ reaches a plateau the system present off-diagonal long range order (ODLRO), the condensate fraction $n_0$ being this asymptotic value $n_0 = \lim_{r\to\infty}\rho_1(r)$. The theoretical study of solid $^4$He at low temperatures cannot be developed analytically via a perturbative approach. It is therefore necessary the use of microscopic approaches to provide a reliable description of this phenomenon. Quantum Monte Carlo (QMC) methods have been extensively used for this purpose, but, so far, they have not been able to reproduce the experimental findings on supersolid $^4$He. Path integral Monte Carlo simulations (PIMC) results have shown that a commensurate perfect crystal does not exhibit either superfluid fraction\cite{Ceperley04} or ODLRO\cite{Ceperley06}. Instead, a non-zero condensate fraction has been observed in simulations of crystals presenting defects, such as vacancies\cite{Galli06} or grain boundaries\cite{Pollet07}, and in simulations of an amorphous state of the solid\cite{Boninsegni06}. Nevertheless, these simulations are not able to provide results in complete agreement with the experimental ones and therefore does not bring to a definitive answer of the supersolidity problem.

In this paper, we perform PIMC calculations of the one-body density matrix and of the momentum distribution in solid $^4$He by means of a very accurate sampling scheme. We will study commensurate and incommensurate hcp crystals, both at zero and finite temperature. In section \ref{PIMC}, we describe the methods used in our simulations. In section \ref{Results}, we present the results obtained, at first, for perfect hcp crystals at different densities, and subsequently, for hcp crystals presenting vacancies. Finally, our conclusions are comprised in section \ref{conclusions}.

\section{The Path Integral Monte Carlo method}\label{PIMC}
The path integral Monte Carlo (PIMC) method provides a fundamental approach in the study of strongly interacting quantum system at finite temperature\cite{CeperleyRev}. It is well known that the partition function
\begin{equation}\label{PartitionFunction}
Z = \rm{Tr}(e^{-\beta\hat{H}}) = \int dR \langle R \vert e^{-\beta\hat{H}} \vert R \rangle
\end{equation}
allows for a full microscopic description of the properties of a given system with Hamiltonian $\hat{H} = \hat{K} + \hat{V}$ at a temperature $T = (k_B\beta)^{-1}$ (the complete basis we use is the position basis $\vert R \rangle = \vert {\bf r}_1, \ldots {\bf r}_N \rangle$ where the $N$ particles are labeled). The noncommutativity of the kinetic  and the potential energy operators (respectively, $\hat{K}$ and $\hat{V}$) makes impractical a direct calculation of $Z$ from Eq. \ref{PartitionFunction}.

The basic idea of PIMC is to use the convolution property of the thermal density matrix $\rho(R,R';\beta)=\langle R \vert e^{-\beta\hat{H}} \vert R' \rangle$, in order to rewrite the partition function as
\begin{equation}\label{Convolution}
Z = \int \prod_{i=0}^{M-1}dR_i \, \rho(R_i,R_{i+1};\varepsilon) \ ,
\end{equation}
with $\varepsilon=\beta/M$ and the boundary condition $R_M = R_0$. For sufficiently large $M$, we recover the high-temperature limit for the thermal density matrix, where it is easy to separate the kinetic contribution from the potential one (Primitive Approximation). If one ignores the quantum statistics of the particles, the distribution law appearing in Eq. \ref{Convolution} is positive definite and can be interpreted as a probability distribution function which can be sampled by standard metropolis Monte Carlo methods.

In practice, the PIMC method consists in mapping the finite-temperature quantum system to a classical system made up of closed ring polymers. This technique may be referred as an "exact" method, in the sense that using an accurate approximation for the high-temperature density matrices, the results are not affected by this approximation within the statistical error. However, its disadvantage is that the number $M$ of convolution terms (beads) necessary to reach the convergence of Eq. \ref{Convolution} to the exact value of $Z$ is inversely proportional to the temperature of the system: this means that, when approaching the interesting quantum regime at very low temperature, $M$ increases fast making simulations hard, if not impossible, due to the very low efficiency in the sampling of the long chains involved.

To overcome this problem, it is important to develop high-order approximation schemes for the density matrix, able to work with larger values of $\varepsilon$. The approximation we use in this work is called Chin Approximation (CA)\cite{Sakkos09}. CA is based on a fourth order expansion of the $e^{-\beta\hat{H}}$ which makes use of the double commutator $[[\hat{V},\hat{K}],\hat{V}]$, this term being related to the gradient of the interatomic potential. With respect to Takahashi-Imada Approximation\cite{TakahashiImada}, which is accurate to fourth order only for the trace, the new feature appearing in the CA is the presence of coefficients weighting the different terms in the expansion of the action: these coefficients are continuously tunable, making possible to force the error terms of fourth order to roughly cancel each other and get an effective sixth-order approximation.

An additional problem we have to deal with when simulating quantum many-body systems with PIMC arises from the indistinguishable nature of the particles. If we deal with bosons like $^4$He, the indistinguishability of particles does not change the positivity of the probability distribution in Eq. \ref{Convolution} and the symmetry of $\rho(R,R';\beta)$ can be recovered via the direct sampling of permutations between the ring polymers representing the quantum particles. To this purpose, a very efficient sampling is performed by the Worm Algorithm (WA)\cite{BoninsegniWorm}: the basic idea of this technique is to work in an extended configuration space, given by the union of the ensemble $Z$, formed by the usual closed-ring configurations, and the ensemble $G$, which is made up of configurations where all the polymers but one are closed. Thanks to the presence of an open polymer, we are able to search the atoms involved in a permutation cycle by means of single particle updates, which do not suffer of a low acceptance rate and guarantee an efficient and ergodic sampling of the bosonic permutations. We have to notice that the probability distribution used to sample the configurations in $G$ is not equal to the one appearing in Eq. \ref{Convolution} and, therefore, these configurations cannot be used to calculate diagonal properties, such as the energy or the superfluid density. However, the $G$-configurations can be used to compute off-diagonal observables such as the one-body density matrix $\rho_1({\bf r}_1,{\bf r}_1')$. Furthermore, the WA being able to sample both diagonal and off-diagonal configuration, is able to give an estimation of the normalization factor of $\rho_1({\bf r}_1,{\bf r}_1')$. In this way, we are able to compute the properly normalized one-body density matrix and so to avoid the systematical uncertainties introduced by a posteriori normalization factor.

The PIMC technique can be extended to zero temperature, in the so called path integral Ground State (PIGS) method\cite{SarsaPIGS}. Indeed, we can see that the same imaginary-time evolution operator appearing in the definition of the thermal density matrix, can be used to project a trial wave function $\Psi_T$ to the exact ground-state wave function $\Psi_0$ according to the formula
\begin{equation}\label{PIGS_wf}
\Psi_0(R) = \lim_{\beta \to \infty} \int dR' \rho(R,R';\beta) \Psi_T(R') \ .
\end{equation}
As in the finite temperature approach, one can compute the ground-state averages of physical observables by factorizing $\rho(R,R';\beta)$ and studying the convergence with the number $M$ of convolution terms. A good approximation for $\rho(R,R';\varepsilon)$ at small $\varepsilon$, like CA, makes possible to choose $\Psi_T = 1$ in Eq. \ref{PIGS_wf} and thus to obtain exact and completely model-independent results still with a very small number of beads $M$\cite{Rota10}.

\section{Results}\label{Results}
\begin{figure}
\begin{center}
\includegraphics[width=0.38\linewidth,angle=270,keepaspectratio]{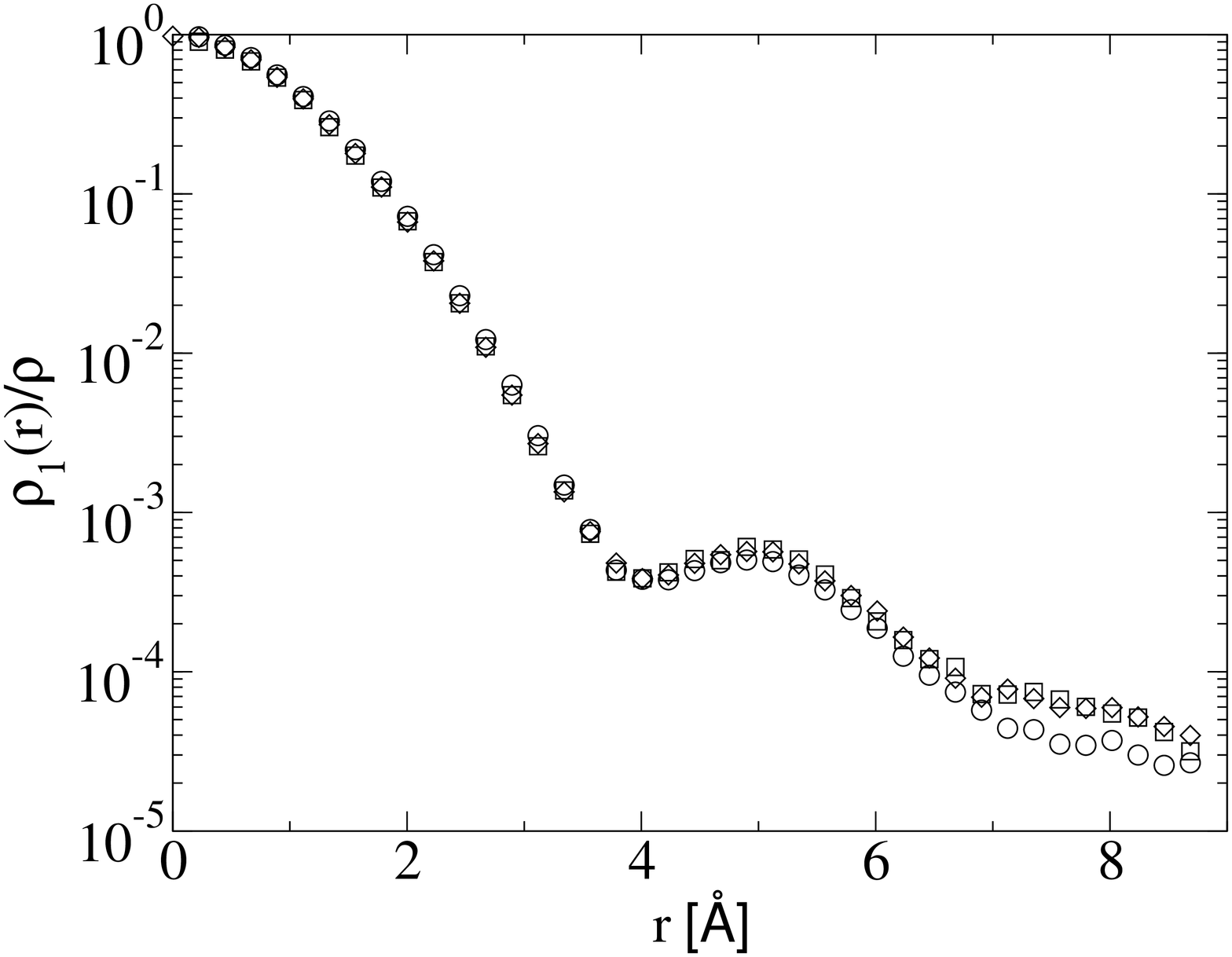}
\includegraphics[width=0.38\linewidth,angle=270,keepaspectratio]{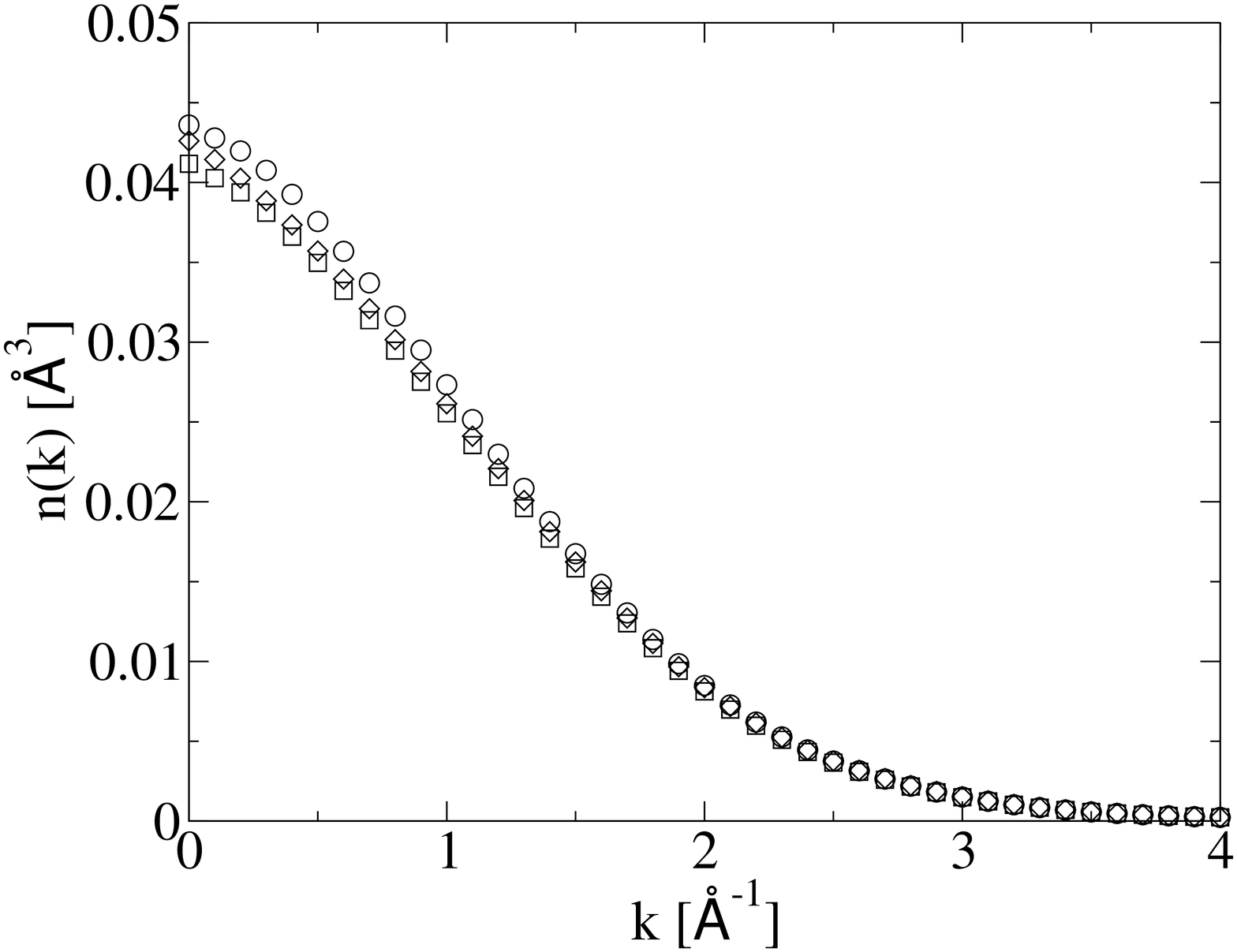}
\end{center}
\caption{The one-body density matrix $\rho_1(r)$ (left) and the momentum distribution $n(k)$ (right) for a commensurate hcp crystal at density $\rho = 0.0294 \, \rm{\AA^{-3}}$ and at different temperatures: $T = 0 \, {\rm K}$ (circles), $T = 1 \, {\rm K}$ (squares) and $T = 2 \, {\rm K}$ (diamonds). Statistical errors are below symbol size.}\label{rho1nk0491sigma-3}
\end{figure}

We have carried out PIMC simulations of solid $^4$He using a perfect hcp lattice with a simulation box containing $N = 180$ atoms interacting with an Aziz pair potential\cite{Aziz}. At first, we perform calculations for a solid at a density $\rho = 0.0294 \, {\rm \AA^{-3}}$, at three different temperatures. The results for $\rho_1(r)$ and $n(k)$ are shown in Figure \ref{rho1nk0491sigma-3}. We can see that, at large $r$, $\rho_1(r)$ decays exponentially, indicating that this system does not present ODLRO. We can also see that there is not a dependence of the momentum distribution with the temperature, in agreement with previous results by Clark and Ceperley\cite{Ceperley06}.

In order to compare our results to the ones obtained by neutron scattering experiments, we compute the Compton profile of the longitudinal momentum distribution $J(y)$, which is $n(\bf{k})$ projected along the direction of the momentum $\bf{Q}$ of the incoming neutron and it is of easier experimental access than $n({\bf k})$. In the Impulse Approximation, which describe well the inelastic neutron scattering at high momentum transfer, $J(y)$ and $n(k)$ are related by the formula\cite{GlydeBook}
\begin{equation}
J(y) = \int d{\bf k} \, n({\bf k}) \, \delta(y - k_Q) =  2\pi \int_{|y|}^\infty dk \, k \, n(k)
\end{equation}
being $k_Q = {\bf k} \cdot \frac{\bf Q}{|Q|}$.

\begin{figure}
\begin{center}
\includegraphics[width=0.38\linewidth,angle=270,keepaspectratio]{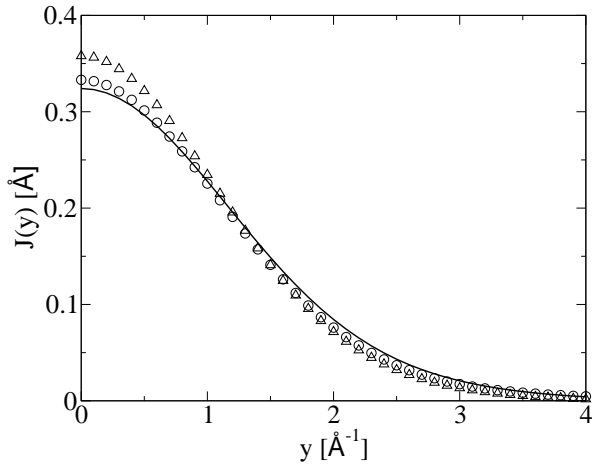}
\end{center}
\caption{The longitudinal momentum distribution $J(y)$: our results at $T = 0 \, \rm{K}$ and $\rho = 0.0294 \, {\rm \AA^{-3}}$ (circles) are compared with the fit obtained by Diallo {\it et al.}\cite{Diallo07} from neutron scattering experiment at $T = 80 \, \rm{nK}$ and $\rho = 0.0301 \, \rm{\AA^{-3}}$ (solid line) and with the PIMC results of Ceperley\cite{Ceperley89} for a bcc solid at $T = 1.67 \, \rm{K}$ and $\rho = 0.0288 \, \rm{\AA^{-3}}$.}\label{LongitudinalDistribution}
\end{figure}

In Figure \ref{LongitudinalDistribution}, we compare the results for the longitudinal momentum distribution obtained from our $n(k)$ at zero temperature with the fit obtained from experimental measurement by Diallo {\it et al.} for the same quantity in solid $^4$He at molar volume $V_m = 20.01 \, \rm{cm^3/mol}$ ($\rho = 0.0301 \, \rm{\AA^{-3}}$) and a temperature $T = 80 \, \rm{nK}$\cite{Diallo07}. We can see that our result are in a good agreement with the experimental ones.

In Figure \ref{LongitudinalDistribution}, we have also plotted the $J(y)$ obtained by Ceperley in a PIMC simulation for a bcc crystal at a density close to the melting of solid $^4$He $\rho = 0.0288 \, \rm{\AA^{-3}}$  and at a temperature $T = 1.67 \, \rm{K}$\cite{Ceperley89}. The difference between our momentum distribution and the one computed by Ceperley has to be attributed to the larger density of the crystal we are simulating, and indicates that a larger coordination between the atoms in the solid cause a depletion of the low momentum states.

\begin{figure}
\begin{center}
\includegraphics[width=0.38\linewidth,angle=270,keepaspectratio]{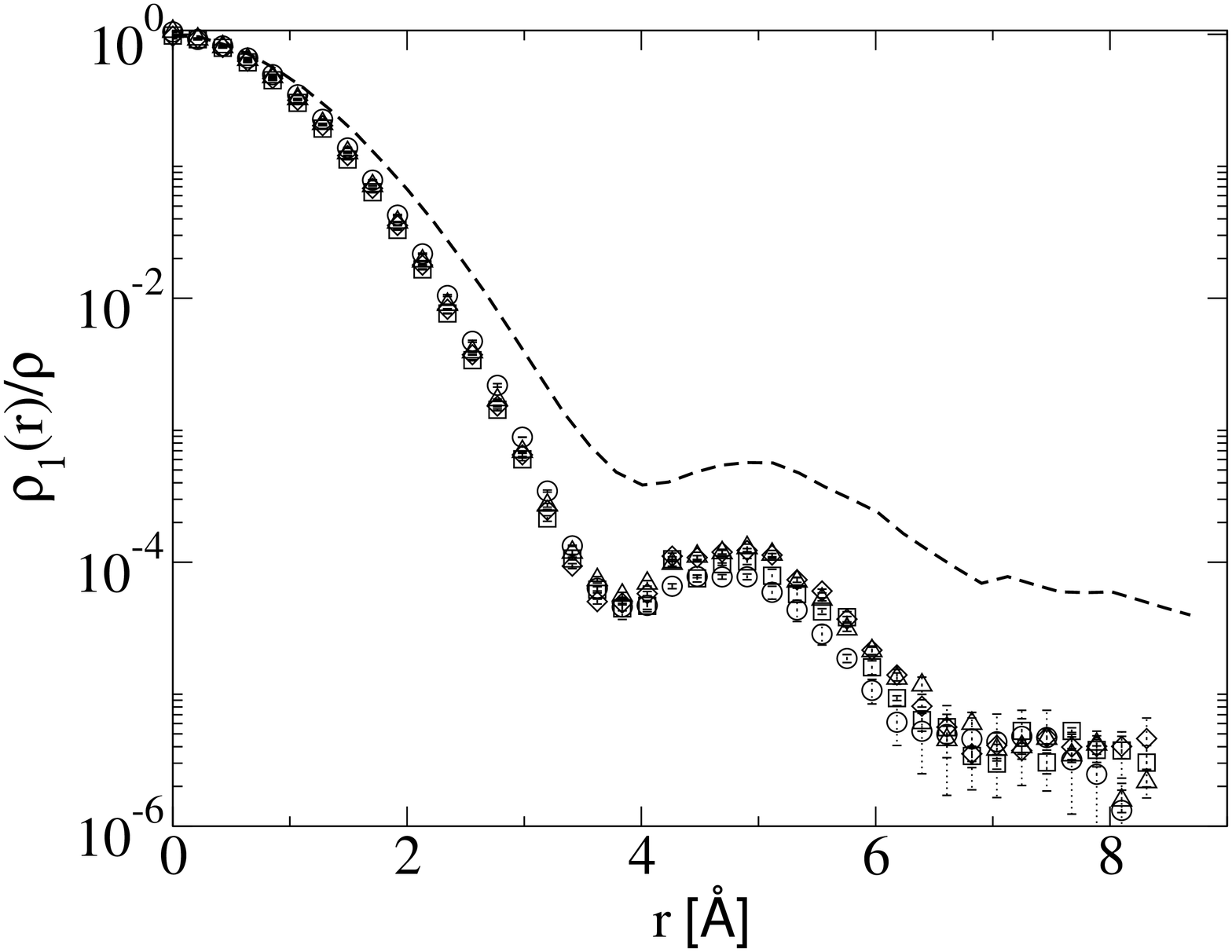}
\includegraphics[width=0.38\linewidth,angle=270,keepaspectratio]{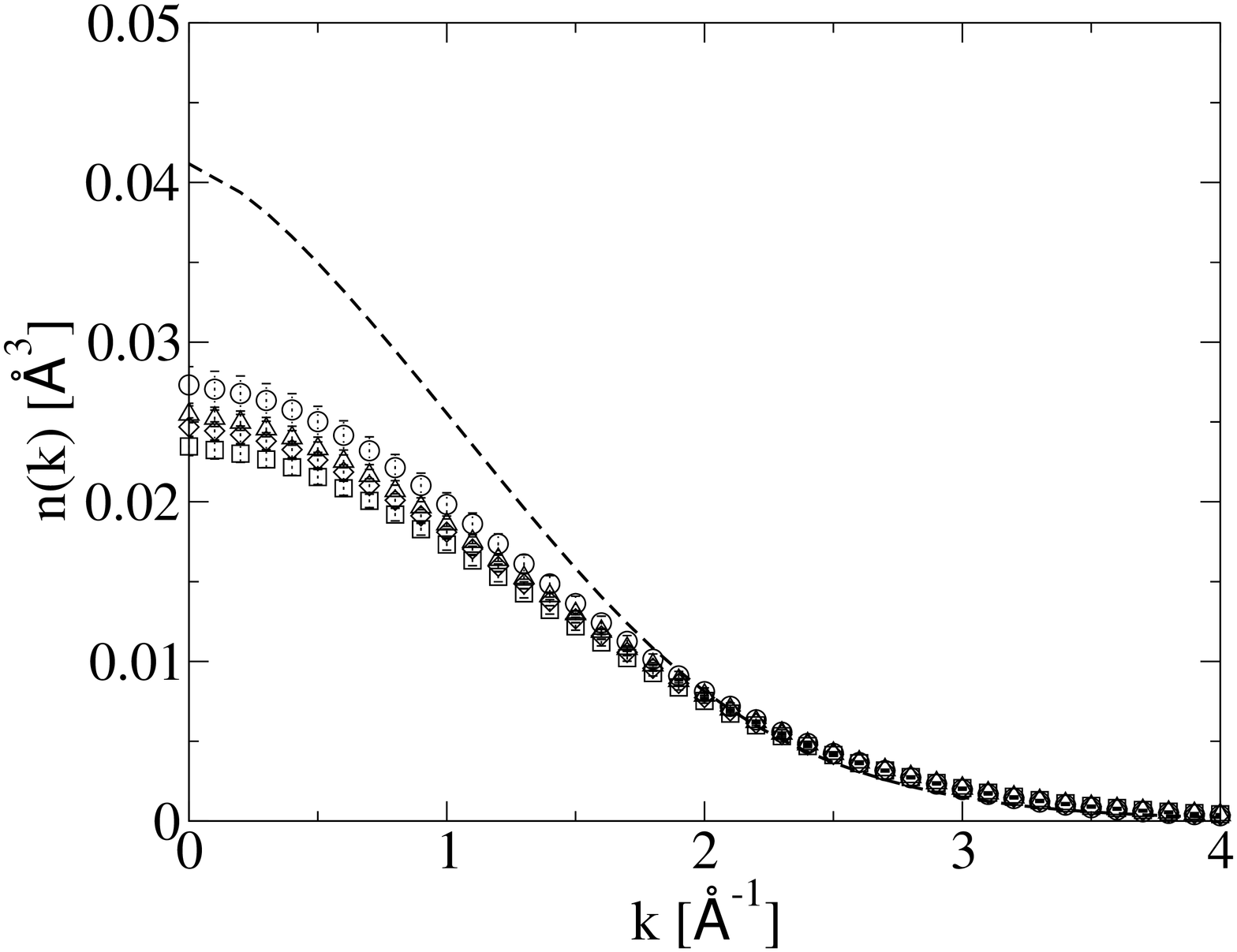}
\end{center}
\caption{The one-body density matrix $\rho_1(r)$ (left) and the momentum distribution $n(k)$ (right) for a commensurate hcp crystal at density $\rho = 0.0335 \, \rm{\AA^{-3}}$ and at different temperatures: $T = 0 \, {\rm K}$ (circles), $T = 1 \, {\rm K}$ (squares), $T = 2 \, {\rm K}$ (diamonds) and $T = 3 \, {\rm K}$ (triangles). The dashed lines represent the same quantities computed for hcp crystal at $T = 1 \, {\rm K}$ and $\rho = 0.0294 \, \rm{\AA^{-3}}$}\label{rho1nk056sigma-3}
\end{figure}

This behavior is confirmed when simulating a crystal at higher densities. Figure \ref{rho1nk056sigma-3} shows $\rho_1(r)$ and $n(k)$ in a crystal at the density $\rho = 0.0335 \, \rm{\AA^{-3}}$. For this density, the solid phase is stable over a larger range of temperature, making us able to simulate the system at temperatures up to $T = 3 \, \rm{K}$. Comparing the results for the two different densities, we see that, when $\rho$ increases, the one-body density matrix decays faster to zero and that the occupation of the low momentum states is appreciably decreased. It is also interesting to notice that $\rho_1(r)$ and $n(k)$ are nearly independent of $T$ even for temperatures larger than $T_{\lambda}$, that is the superfluid transition temperature in the liquid phase.

\begin{figure}
\begin{center}
\includegraphics[width=0.38\linewidth,angle=270,keepaspectratio]{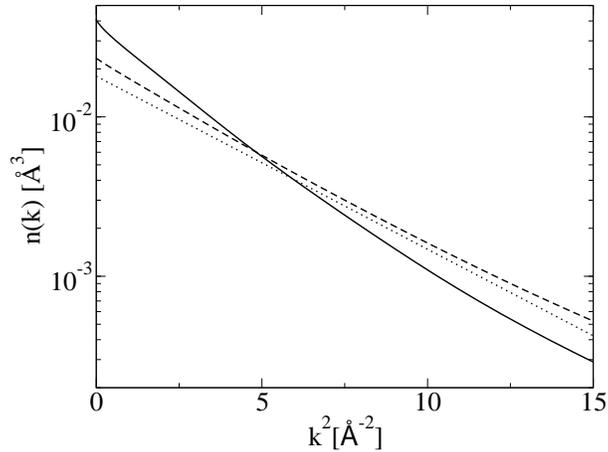}
\end{center}
\caption{The momentum distribution $n(k)$ as a function of $k^2$ at $T = 1 \, {\rm K}$ and densities $\rho = 0.0294 \, \rm{\AA^{-3}}$ (solid line) and $\rho = 0.0335 \, \rm{\AA^{-3}}$ (dashed line). The dotted line represent a Gaussian $n(k)$ and it is used to guide the eye.}\label{CompareGaussian}
\end{figure}

It is important to notice that at both densities, the shape of $n(k)$ differs significantly from the classical Gaussian Maxwell-Boltzmann distribution. This feature is clearly shown in Figure \ref{CompareGaussian}, where we plot, on a logarithmic scale, $n(k)$ as a function of $k^2$. From this graph, it is easy to see the differences between the momentum distributions we obtained from PIMC and the straight line which represent a Gaussian distribution. We also notice that these differences are smaller in the system at $\rho = 0.0335 \rm{\AA^{-3}}$, indicating that solid helium becomes more classic when the density increases.

\begin{figure}
\begin{center}
\includegraphics[width=0.38\linewidth,angle=270,keepaspectratio]{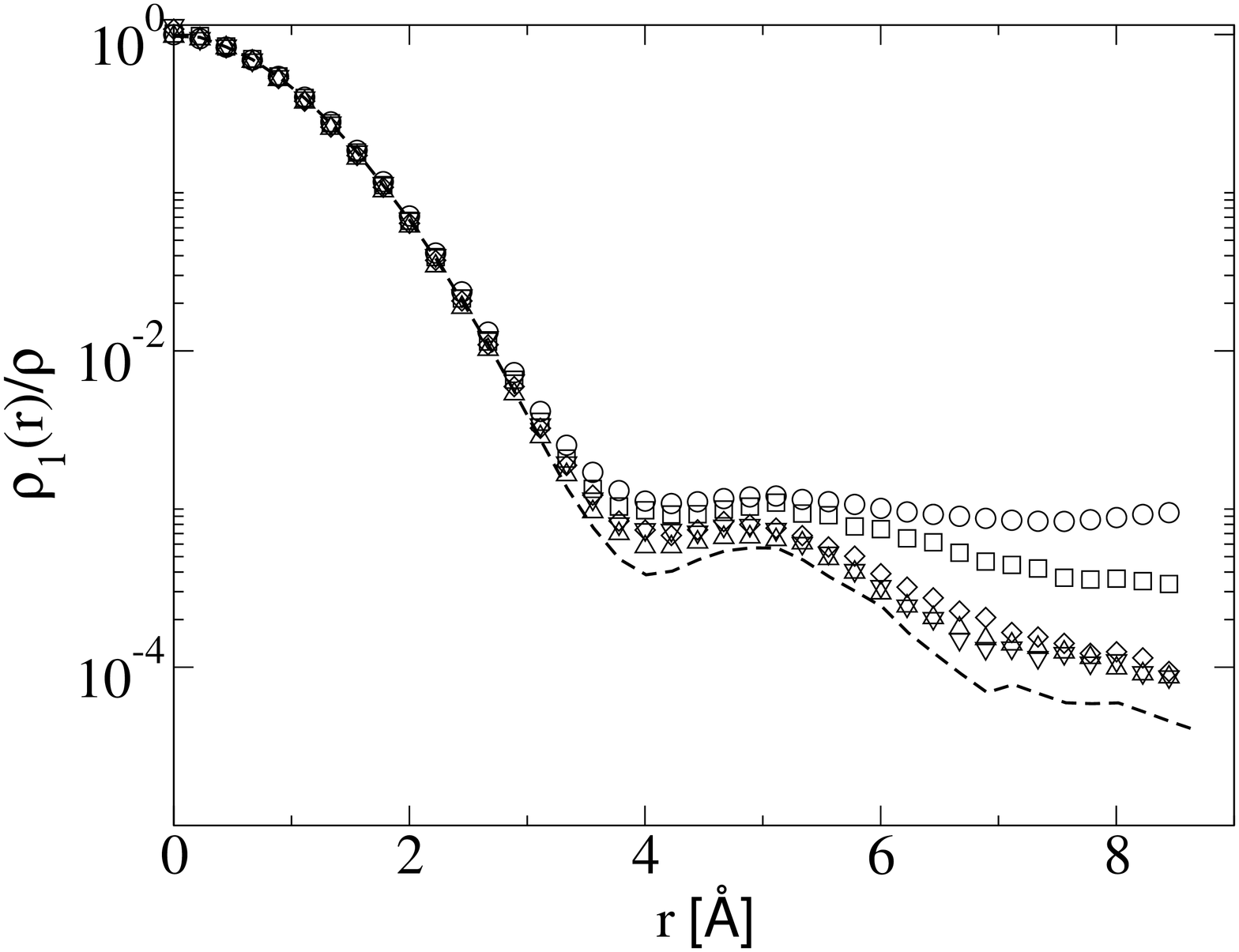}
\includegraphics[width=0.38\linewidth,angle=270,keepaspectratio]{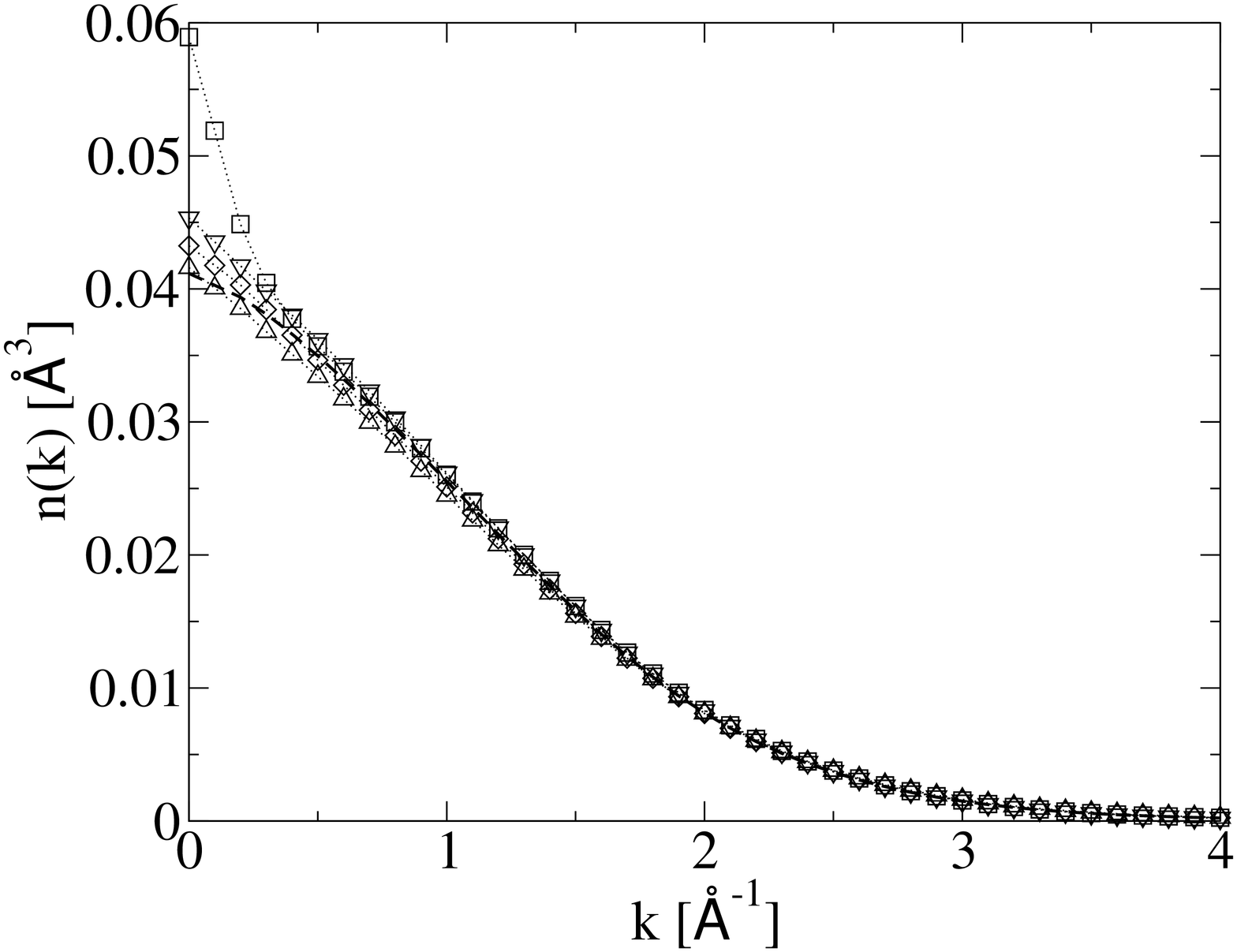}
\end{center}
\caption{The one-body density matrix $\rho_1(r)$ (left) and the momentum distribution $n(k)$ (right) for a hcp crystal presenting a vacancy at density $\rho = 0.0294 {\rm \AA^{-3}}$ and at different temperatures: $T = 0 \, {\rm K}$ (circles), $T = 0.5 \, {\rm K}$ (squares), $T = 0.75 \, {\rm K}$ (diamonds), $T = 1 \, {\rm K}$ (triangles up) and $T = 2 \, {\rm K}$ (triangles down). The dashed lines represent the same quantities computed for a commensurate hcp crystal at $T = 1 \, {\rm K}$ and $\rho = 0.0294 {\rm \AA^{-3}}$. Statistical errors are below symbol size.}\label{rho1nkVacancy}
\end{figure}

Finally, in order to give a deeper insight in the debate about supersolidity, it is important to understand how the momentum distribution changes when vacancies are present inside the crystal. For this purpose, we compute $\rho_1(r)$ and $n(k)$ for a system made up of $N = 179$ atoms in a box which fits an $N_s = 180$ sites hcp lattice. We perform our calculations at the density $\rho = 0.0294 \, \rm{\AA^{-3}}$ and over a range of temperatures from $T = 0 \, \rm{K}$ to $T = 2 \, \rm{K}$. The results presented in figure \ref{rho1nkVacancy} show that $\rho_1(r)$ at $T = 0 \, {\rm K}$ presents a non-zero asymptote at large $r$, indicating that BEC is present inside the system. Our estimation of the condensate fraction is $n_0 = (9.0 \pm 0.8) \times 10^{-4}$, in agreement with Diffusion Monte Carlo results\cite{Cazorla09}.

Instead, at finite temperature, we do not see a clear signal of ODLRO since all the $\rho_1(r)$ decay exponentially at large $r$: this indicates that the onset temperature $T_c$ at which the condensate appears in the system is below the lowest temperature we study, that is $T_c < 0.5 \, \rm{K}$. Nevertheless, from the study at finite temperature we notice an interesting dependence of $\rho_1(r)$ and $n(k)$ with $T$: we can see that, for $T \ge 0.75 \, \rm{K}$, $\rho_1(r)$ differs only slightly from the same quantity calculated for the perfect crystal, while no difference can be seen for $n(k)$ in the two systems. Instead, at $T = 0.5 \, \rm{K}$, we see that the incommensurate crystal behaves differently since an additional peak centered in ${\bf k} = 0$ appears in the momentum distribution. This result may be explained assuming that at temperatures $T \ge 0.75 \, \rm{K}$, the vacancy creates only a local distortion of the lattice which does not affect the momentum distribution, while, below a critical temperature $T_0 < 0.75 \, \rm{K}$, the defect begins to delocalize and to allow a larger occupation of the low momentum states.

\section{Conclusions}\label{conclusions}
To summarize, we have computed the momentum distribution in solid $^4$He by means of PIMC methods. These calculations are of fundamental importance in the study of BEC properties of quantum solids. The use of an effective sixth-order approximation for the action allows us to study the system by means of a very accurate sampling scheme, and the implementation of Worm Algorithm makes possible to calculate one-body density matrices which are correctly normalized. We have seen that our results are in good agreement with the ones obtained in neutron scattering experiments and indicates that solid $^4$He is highly anharmonic, even though its behavior approaches the classical one at high densities. We have also shown that the presence of defects like vacancies affects the momentum distribution only at very low temperatures ($T < 0.75 \, \rm{K}$). Calculations of $\rho_1(r)$ and $n(k)$ in defected crystals at lower temperatures are now under way.

\subsection*{Acknowledgements}
We acknowledge partial financial support from DGI (Spain) grant No. FIS2008-04403 and Generalitat de Catalunya grant No. 2009SGR-1003


\end{document}